# Giant Tunability of Intersubband Transitions and Quantum Hall Quartets in Few-Layer InSe Quantum Wells

Dmitry Shcherbakov[1*], Greyson Voigt[1], Shahriar Memaran[2,3], Gui-Bin Liu[4], Qiyue Wang[5], Kenji Watanabe[6], Takashi Taniguchi[7], Dmitry Smirnov[2], Luis Balicas[2,3], Fan Zhang[5], Chun Ning Lau[1†]

[1] Department of Physics, The Ohio State University, Columbus, OH 43221, USA.
[2] National High Magnetic Field Laboratory, Tallahassee, FL 32310, USA
[3] Department of Physics, Florida State University, Tallahassee, FL 32306, USA
[4] School of Physics, Beijing Institute of Technology, 100081 Beijing, China
[5] Department of Physics, The University of Texas at Dallas, 800 West Campbell Road, Richardson, Texas 75080-3021, USA
[6] Research Center for Electronic and Optical Materials, National Institute for Materials Science, 1-1 Namiki, Tsukuba 305-0044, Japan
[7] Research Center for Materials Nanoarchitectonics, National Institute for Materials Science, 1-1 Namiki, Tsukuba 305-0044, Japan

**Abstract**

**A two-dimensional (2D) quantum electron system is characterized by the quantized energy levels, or subbands, in the out-of-plane direction. Populating higher subbands and controlling the inter-subband transitions have wide technological applications such as optical modulators and quantum cascade lasers. In conventional materials, however, the tunability of intersubband spacing is limited. Here we demonstrate electrostatic population and characterization of the second subband in few-layer InSe quantum wells, with giant tunability of its energy, population, and spin-orbit coupling strength, via the control of not only layer thickness but also out-of-plane displacement field. A modulation of as much as 350% or over 250 meV is achievable, underscoring the promise of InSe for tunable infrared and THz sources, detectors and modulators.**

---

[*] Current address: Department of Physics, Carnegie Mellon University, Pittsburgh, PA 15213. Email: dish@cmu.edu.
[†] Email: lau.232@osu.edu

In a conventional quantum well (QW), charge carriers are confined in a mesoscopic layer whose thickness is comparable to their de Broglie wavelength. The advent of 2D van der Waals (vdW) materials enables a new class of QWs that can be atomically thin, widely tunable by a number of knobs, mechanically flexible and compatible with surface probes. However, to date most electronic and optoelectronic studies of 2D vdW materials focus on the properties of the lowest electronic subband. Recently, there has been increasing interest in exploring inter-subband transitions via infrared nano-imaging [1], photo- and electro-luminescence [2] and resonant tunneling [3, 4] studies. Yet a systematic characterization and *control* of the second subband and the intersubband transition in 2D vdW semiconductors has been lacking to date.

Here we demonstrate electrostatically induced population of the second subband in InSe field effect transistors that are 7-10 layers thick, and extract the effective mass of the charge carriers and the strength of the Rashba spin-orbit coupling (SOC) of the subbands from quantum oscillations, which are supported by our first-principle calculations. For a given thickness $L$, the energetic spacing between the first and second subbands $E_{12}$ scales quadratically with $E_\perp$, with a tunability coefficient that increases with $L$; as $L$ varies, their minimum spacing scales as $1/L^2$. At high magnetic fields, the simultaneous occupation of two subbands, together with the helical spin degrees of freedom, leads to the formation of electronic quartets in the quantum Hall regime, where the ring-shaped crossings between Landau levels from the two subbands reveal a series of quantum phase transitions between paramagnetic and helical magnetic states.

InSe is a layered semiconductor with a layer-dependent band gap, large photoresponsivity [5-9], large gate-tunable Rashba spin-orbit coupling [10-13], high mobility [10, 14-17], and high saturation current [15]. Large-scale synthesis has been demonstrated [12, 18, 19]. In this work, devices are fabricated by encapsulating few-layer InSe sheets between hexagonal BN (hBN) layers, which are etched into Hall bar geometry and coupled to few-layer graphene electrodes [10, 14]. An $Al_2O_3$/Au top gate is deposited on the channel region, and the $SiO_2$/Si substrate serves as a back gate. Here we focus on InSe sheets that are 7 to 10 layers thick, which are sufficiently thick to enable the second subband to be populated via electrostatic gating, but sufficiently thin to function as a single narrow quantum well (as opposed to wide quantum wells that host two separate 2D electron gases on top and bottom surfaces [20]). All data are taken at 300 mK unless otherwise specified.

Fig. 1a-c plots the differentiated longitudinal resistance *dR/dn* (where *n* is the carrier density) from three different regions of a single InSe sheet, which are 6, 7 and 8-layer thick, respectively, as a function of the perpendicular magnetic field $B$ and the charge density $n_{tg}$ induced by the top gate, while the back gate voltage is maintained at 75V. At low densities, we observe prominent Shubnikov de Haas (SdH) oscillations arising from a single Landau fan in all three regions. In the two thicker regions, additional sets of oscillations emerge at higher densities, indicating that the Fermi level reaches the second subband (Fig. 1b-c). The electronic band structures of 7-layer InSe obtained by first-principles calculations, shown in Fig. 1e, indeed feature the presence of subbands and Rashba SOC, in harmony with previous theoretical studies [21-23]. The onset charge density for reaching the second subband, $n_{on}$, is strongly thickness dependent. Fig. 1d plots $n_{on}$ (left axis) for six devices of different thicknesses, and the right axis shows the energetic separation $E_{12}$ between the first two subbands, calculated from $E_{12} = \frac{\hbar^2 (2\pi n_{on})}{2 m_1^*}$, where

$m_1^* = 0.14 m_e$ is the subband's in-plane effective mass ($m_e$ is the rest mass of the electron) [10, 14]. From the particle-in-a-box model, $E_{12} = \frac{3\hbar^2 \pi^2}{2 m_\perp (L+2v)^2}$, where $m_\perp$ is the effective mass in the out-of-plane direction, $\hbar$ the reduced Planck constant, and $v = 1.42$ is a parameter that accounts for the anharmonicity of the bands [21, 22]. From the slope of the fitting, we estimate that $m_\perp = 0.09\ m_e$, in agreement with the bulk value of $0.08 m_e$ [24]. We also note that the measured $E_{12}$ values are smaller by a factor of ~2 than those from the first-principle calculations; this discrepancy may arise from the anharmonicity of the bands, and/or neglecting the electronic interactions filling the first subband in our first-principles calculations.

To further characterize the second subband, we measure $R(n, B)$, where $n = \frac{C_{Tg} V_{Tg} + C_{Bg} V_{Bg}}{2e}$ is the total charge density, while keeping the perpendicular displacement field $E_\perp = \frac{C_{Tg} V_{Tg} - C_{Bg} V_{Bg}}{2\varepsilon_0}$ zero (Fig. 2a). Here, $C_{Tg}$ and $C_{Bg}$ are the capacitances per unit area between the two gates and InSe, $e$ is the electron charge, and $\varepsilon_0$ is the permittivity of vacuum. A prominent feature revealed by the high mobility sample is the abrupt change in the slopes of the Landau fan originating from the first band at $n_{on}$, as outlined by the dotted line in Fig. 2a. This change in slope reflects the onset of multi-band transport: the total charge density is now divided between two bands, in proportion to their respective density of states (DOS) $\frac{m_i^*}{\pi \hbar^2}$, where $m_i^*$ is the in-plane effective mass of the $i$-th subband. Within the effective mass approximation, the charge densities residing in the two subbands are $n_1 = (n - n_{on}) \frac{m_1^*}{m_1^* + m_2^*} + n_{on}$ and $n_2 = (n - n_{on}) \frac{m_2^*}{m_1^* + m_2^*}$. Thus, the Landau fan from the first subband experiences a slope enhancement by a factor of $\frac{m_1^* + m_2^*}{m_1^*}$. It follows that from the ratio of the slopes, we can extract the ratio of the effective masses, which is estimated to be $m_2^*/m_1^* \sim 1.9$; using $m_1^* = 0.14 m_e$, as determined in prior reports [10, 14] and first principles calculations, we obtain that $m_2^* \sim 0.27 m_e$, which is significantly enhanced with respect to that of the first subband. This enhancement is qualitatively consistent with the first-principles calculations in this and prior works [21, 22], and should provide further constraints for fine-tuning parameters in band structure calculations of few-layer InSe.

A distinguishing characteristic of few-layer InSe is its large tunable Rashba SOC strength. as we have demonstrated previously [10]. To explore the SOC in the second subband, we examine the Landau fan $R(n_{tg}, B)$ of a 9-layer device that displays prominent beating in the SdH oscillations in both subbands (Fig. 2b). Here $n_{tg}$ is the total charge density induced by top gate voltages, while the back gate is maintained at 0 V. Such beating arises from the interference of the cyclotron orbits of the inner and outer helical bands with different cross-sectional areas of the Fermi surfaces [10, 25, 26]. Fig. 2c plots the Fourier transform of the data, showcasing the two distinct frequencies for both subbands. Their Rashba SOC strengths can be measured from the beating patterns [27], $\alpha \approx \frac{\hbar^2}{m^*} \sqrt{\frac{\pi}{2}} \frac{\Delta n}{\sqrt{n_{avg}}}$, where $\Delta n = (e/h) B_{F,beat}$ is the density difference between the electrons in the outer and inner helical bands, $n_{avg}$ is their average, and $B_{F,beat}$ is the beating frequency in the SdH oscillations. Using Eq. (1), we extract the values of $\alpha$, which vary as a function of $n_{tg}$. The variations in $\alpha$ arises from the varying displacement field $E_\perp$. Fig. 2d plots $\alpha(E_\perp)$ for both subbands, demonstrating that the Rashba SOC of the second subband exhibits a much stronger dependence on $E_\perp$ than that of the first. The magnitude and trend of $\alpha$ is in approximate agreement with DFT calculations [21]. To the best of our knowledge, this is the first demonstration of electronic filling of the second subband in a 2D vdW semiconductor, as well as the first measurement of its effective mass and tunable Rashba SOC.

To illustrate tunability of the inter-subband transition, we first examine the background-subtracted $R(n_{tg}, B)$ data from a 10-layer device, which are taken with $V_{bg}$ maintained at 78V and 47V, respectively, hence at varying $E_\perp$ (Fig. 3a-b). The onset of the second subband, indicated by the arrows, occurs at noticeably different densities. To better characterize the electric control of the onset density's dependence on $E_\perp$, we measure the background-subtracted $R$ as a function of both $n$ and $E_\perp$ for an 8-layer device in the quantum Hall (QH) regime at a constant $B$ (Fig.s 3c-d). Below $n_{on}$, the horizontal stripes correspond to the spin-degenerate QH states from the first subband. With the onset of the second subband, which occurs at $n$ = 4.8 and 7.2x10$^{12}$ cm$^{-2}$ at $B$ = 7.5T and 5T, respectively, an intricate pattern emerges, beyond the simple crossings between LLs arising from the two subbands. Specifically, the otherwise horizontal stripes that originate from the first subband curve towards higher densities, owing to the smaller charge density that reside in the first subband under large $E_\perp$. The curvature is symmetric with respect to $E_\perp$ = -0.07 V/nm, which is the value needed to compensate for the intrinsic inversion asymmetry of the material. The onset of the curvature does not occur at a constant density or displacement field; rather, it occurs at larger $E_\perp$ for higher $n$, as indicated by the green stars in Fig. 3d. As this onset density corresponds to the energetic separation $E_{12}$ between the two subbands, this explicitly demonstrates that $E_{12}$ increases with $E_\perp$.

To quantify the dependence of $E_{12}$ on $E_\perp$, we model $E_{12}$ (in meV) = $a|E_\perp - E_{\perp 0}|^2$ and calculate the DOS from the two subbands as a function of $n$ and $E_\perp$ at a constant magnetic field. Here $a$ is a tunability coefficient in meV/(V/nm)$^2$, $E_\perp$ is in V/nm, $E_{12}$ is in meV, and $E_{\perp 0}$ is the displacement field that minimizes $E_{12}$; note that $E_{\perp 0} \neq 0$ due to the broken structural inversion symmetry of InSe. Using $a \approx 85$, we calculate the DOS from the two subbands as a function of $n$ and $E_\perp$ at a constant magnetic field. The simulation of the 8-layer device at $B$ = 5T is shown in Fig. 3e, which nicely reproduces the experimental data in Fig. 3c. Repeating the same simulation, we find that $a \approx 140$ and 210 for 9-layer and 10-layer devices, respectively. Our first-principles calculations also show a similar trend of displacement field dependent $E_{12}$ and its enhancement with thickness. To better appreciate the magnitude of this tuning of subband separations, we plot the magnitude of $E_{12}$ vs $E_\perp$ for 8-, 9-, and 10-layer devices in Fig. 3f; the percentage changes normalized to their respective minima are shown in Fig. 3g. The dependence of the tunability coefficient $a$ on layer number is shown in Fig. 3f inset. Evidently, the $E_\perp$-controlled tuning effect is dramatic, ranging from 30% at $E_\perp$ = 0.5 V/nm for the 8-layer device, to over 350% or 250 meV at $E_\perp$ = 1 V/nm for the 10-layer device. Such giant tunability is hitherto unobserved, and orders of magnitude higher than any prior reports in 2D electron systems.

Lastly, we explore the LL crossings between the first and second subbands under high magnetic fields, where the interplay between the subband and spin degrees of freedom gives rise to a sequence of field-controllable electronic quartets. Fig. 4a displays the background-subtracted $R(n_{tg}, B)$ of a 10-layer device under $B$ up to 40T. Similar to Fig.s 1-3, two sets of Landau levels are observed. The crossings between LLs from the first subband and the first LL from the second subband form a sawtooth-like structure, similar to those observed in multilayer graphene [28, 29] and GaAs systems [30, 31], evolving into ring-like patterns at higher magnetic field. These "rings" arise from a mechanism analogous to quantum Hall ferromagnetism, when electrons transfer between LLs to form a spin-polarized or spin-helical QH state that minimizes their total energy. This effect is illustrated on Fig. 4b. In regime $S$, the $|+,1\rangle$ and $|-,1\rangle$ LLs of the 1st subband are filled, whereas in regime $P$ the $|+,2\rangle$ and $|-,2\rangle$ LLs of the 2nd subband are filled. Here +/- refer to the inner and outer helical bands split by the combined effects of Rashba SOC and Zeeman. In region $Q$ the $|+,1\rangle$ and $|+,2\rangle$ LLs are filled, and the electron-electron interactions enlarge and distort this regime, where the exchange energy becomes comparable to the energy difference between $|+,2\rangle$ and $|-,1\rangle$ or $|-,2\rangle$ and $|+,1\rangle$ LLs at single-particle level.

Interestingly, we can switch between the subband-polarized and spin-helical QH states solely by varying $E_\perp$ (Fig. 4c). The red-highlighted (green-highlighted) areas in dashed outlines correspond to $\nu=1$ ($\nu=N$) of the second (first) subband, and are subband-polarized. By varying electric field, we can switch into and out of the spin-helical QH states, hence performing purely electrical control of spin texture. We also note that in this QH ferromagnet, the Rashba SOC strength can exceed Zeeman energy, which may enable the realization of a helical state and half-skyrmion defects at an odd filling factor [32].

In conclusion, we demonstrated population of the first and second electronic subbands in few-layer InSe quantum wells, and determined the effective mass and the Rashba SOC strengths of both subbands. In a high magnetic field, crossings between Landau levels arising from these two subbands give rise to quantum Hall quartets, which are distorted by electron-electron interactions and can be tuned by $B$, $n$ and $E_\perp$. Importantly, we demonstrate via an all-electrical means that the energetic separations and interband transitions between the first and the second subband can be modulated by an unprecedented extent, that is more than 350% in a 10-layer device. Such giant gate-tunable intersubband transitions can find wide applications in electronic and optoelectronic technologies.

**Acknowledgement**
DS was supported by NSF/DMR 2128945, GV and CNL are supported by NSF/DMR 2219048. LB is supported by the NSF/DMR 2219003. A portion of this work was performed at the National High Magnetic Field Laboratory, which is supported by the National Science Foundation through NSF/DMR-1644779 and the State of Florida. K.W. and T.T. acknowledge support from the JSPS KAKENHI (Grant Numbers 20H00354, 21H05233 and 23H02052) and World Premier International Research Center Initiative (WPI), MEXT, Japan. Q.W. and F.Z. were supported by the National Science Foundation under grant numbers DMR-1945351 through the CAREER program and DMR-2105139 through the CMP program.

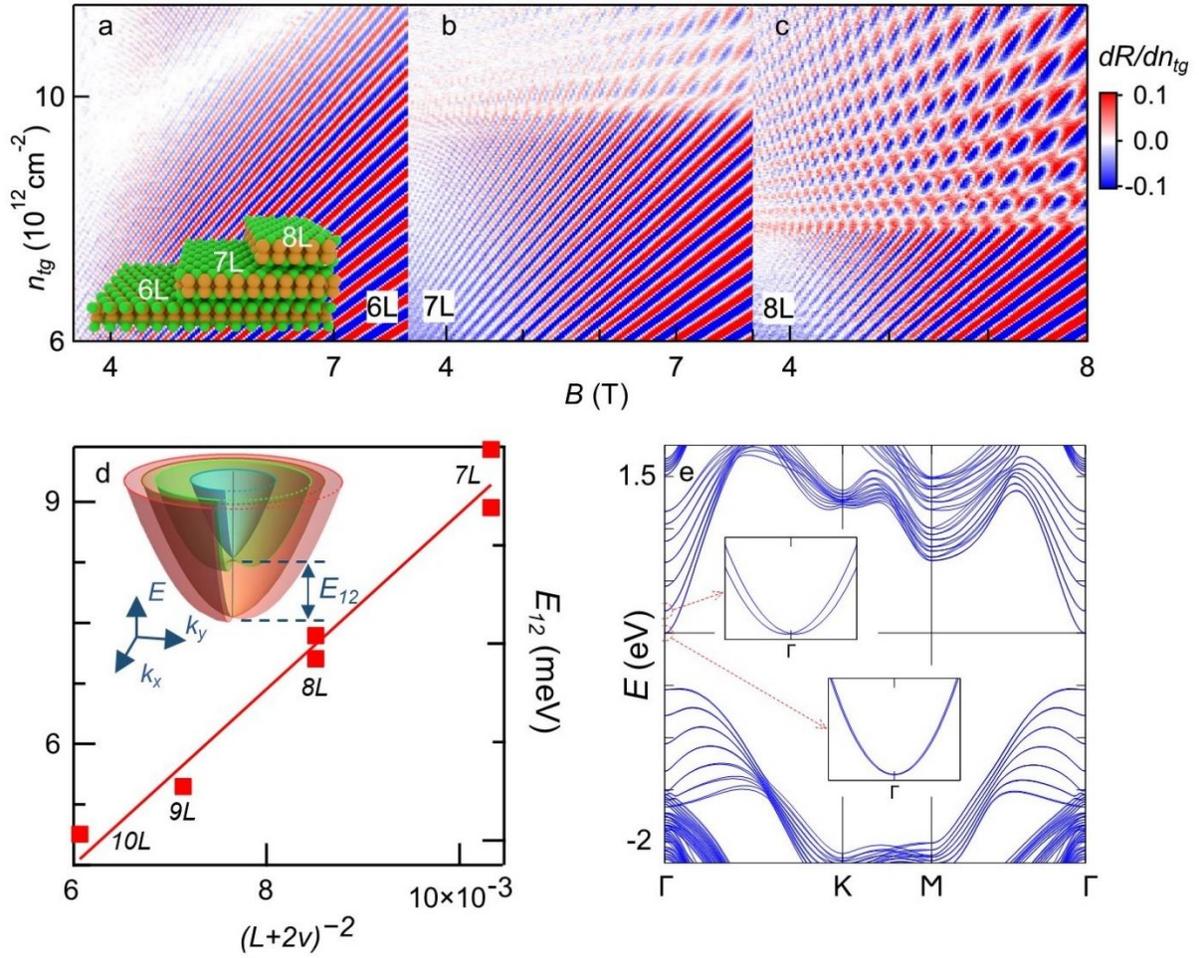

Figure 1. Magnetotransport data from InSe devices as a function of sample thickness. (a-c). $dR/dn_{tg}$ vs $n_{tg}$ and $B$ from a single InSe sheet with regions that are 6-, 7-, and 8-layers thick. Inset: schematics of the flake. (d). Measured onset density for the second subband $n_{on}$ as a function of the effective thickness squared ($L$ is the layer number and $v$=1.42 is a parameter to account for band anharmonicity). The right axis plots the intersubband transition energy $E_{12}$ calculated assuming a constant first subband $m^*$=0.14 $m_e$. The line is a linear fit to the data points. Inset: Illustration of the subbands. (e). Band structure of 7-layer InSe with Rashba SOC obtained by first-principles calculations.

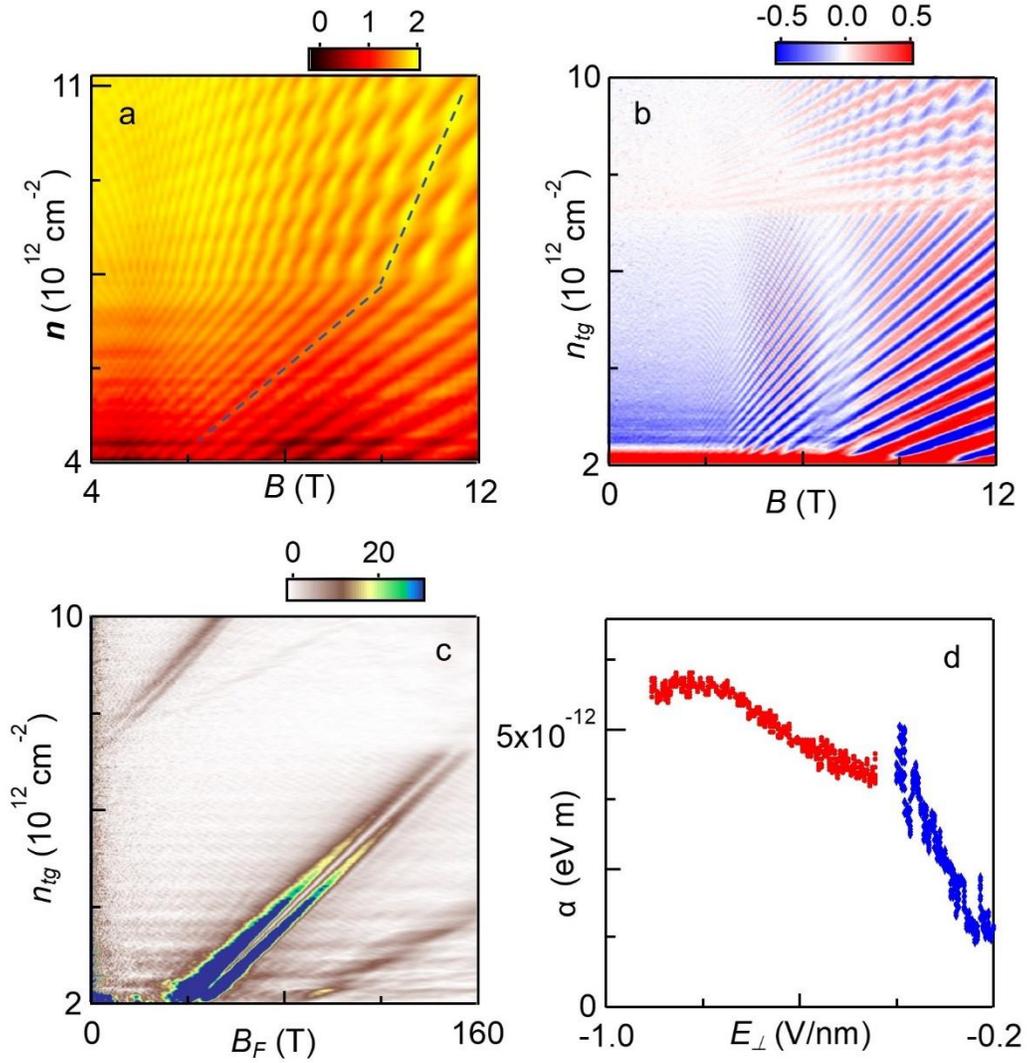

Figure 2. Measurements of the second subband's effective mass and Rashba SOC parameter. (a). $R(n, B)$ for an 8-layer device taken at $E_\perp=0$. The dotted line indicates the change in the slope of the quantum oscillations from the first subband after reaching the onset of the second subband. (b). $dR/dn$ ($n_{tg}$, $B$) for a 9-layer device taken at $V_{bg}$=80V. (c). FFT of the oscillations in (b). The twin peaks arise from the spin-split bands due to the Rashba SOC. (d). Extracted SOC parameter $\alpha$ vs $E_\perp$ for the first (red) and second (blue) subband.

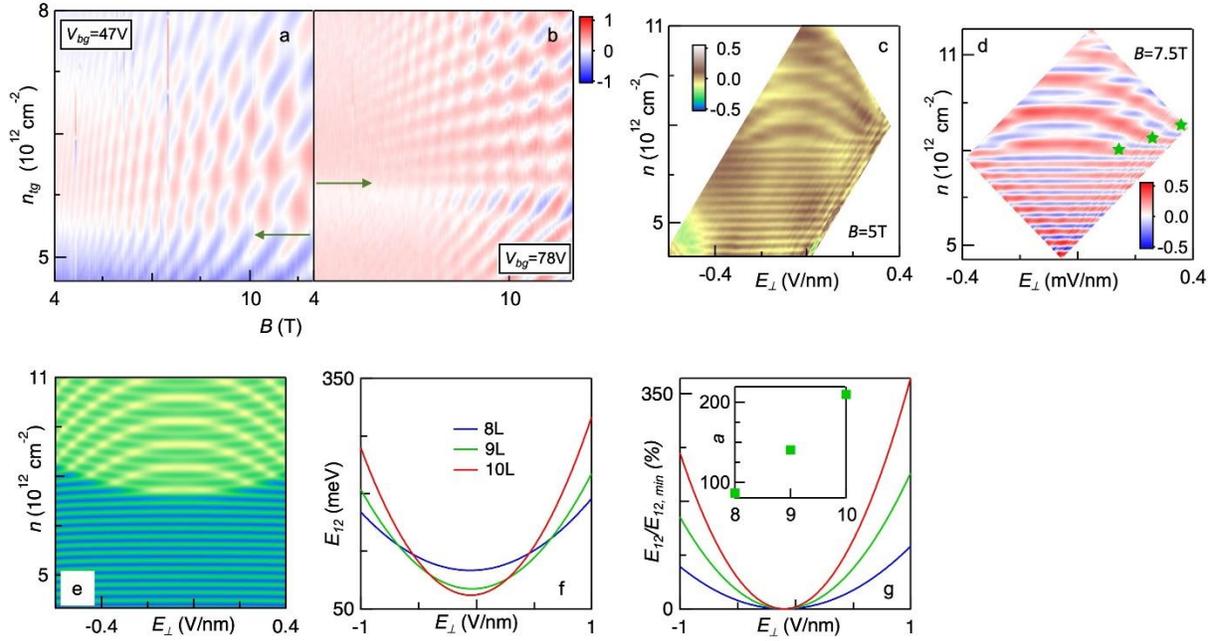

Figure 3. Tuning the intersubband energy with a transverse electric field $E_\perp$. (a-b). Background subtracted $R(n_{tg},B)$ for a 10-layer device, with $V_{bg}$ at 47 and 78V, respectively. The arrows indicate onset charge densities for the second subband. (c-d). Background-subtracted $R(n, E_\perp)$ of an 8-layer device at $B$=5T and 7.5 T, respectively. (e). Simulation of the density of states as a function of $n$ and $E_\perp$ at $B$=5T for the 8-layer device, calculated using $E_{12}$ (in meV)$=aE_\perp^2$, where $a$=85 is the tunability parameter, and $E_\perp$ is in V/nm. (f-g). Extracted intersubband spacing $E_{12}$ and its percentage change vs $E_\perp$ for 8-layer (blue), 9-layer (green) and 10-layer (red) devices, respectively. Inset in g: the tunability parameter $a$ as a function of layer number.

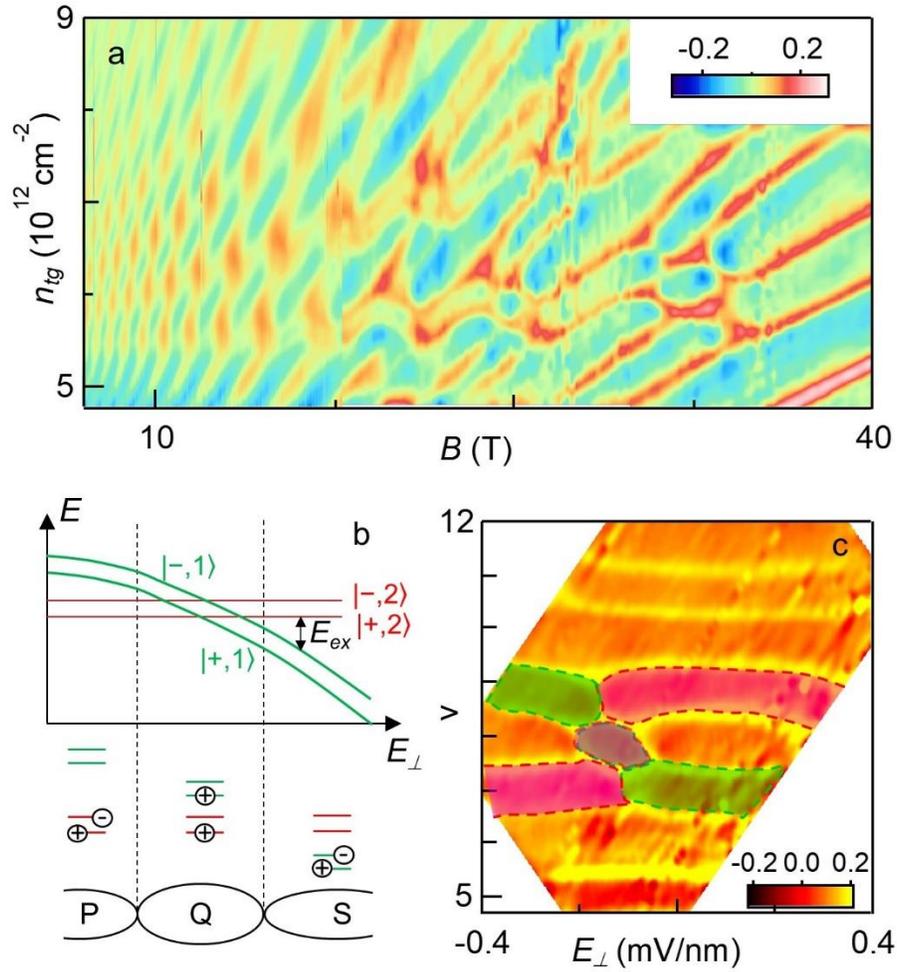

Figure 4. Quantum Hall helical magnetism in InSe under high magnetic fields. (a). Background-subtracted $R(n_{tg}, B)$ for a 10-layer device subjected to high magnetic fields. (b). Schematic of LL crossing and formation of spin-helical quantum Hall state. (c). Background-subtracted $R(v, E_\perp)$ at $B$=34.5T, showing tuning of the QH helical-magnetic states by $E_\perp$.